\documentclass[amsmath,amssymb,column,superscriptaddress]{revtex4}

\usepackage{graphicx}%
\usepackage{dcolumn}%
\def\be{\begin{equation}}
\def\ee{\end{equation}}
\def\beq{\begin{eqnarray}}
\def\eeq{\end{eqnarray}}

\usepackage{bm}
\usepackage{graphicx}
\usepackage{dcolumn}
\usepackage{amsmath}
\usepackage{graphicx, psfrag}
\usepackage{amssymb}
\usepackage[colorlinks=true, citecolor=blue, urlcolor = blue, linkcolor= red, bookmarks=true]{hyperref}
\usepackage{float}
\usepackage{mathtools}
\usepackage{amsmath}
\usepackage{amsfonts}
\usepackage{dcolumn}
\usepackage{hyperref}
\usepackage{subfigure}
\usepackage{pgfplots}
\usepackage{epstopdf}
\usepackage{booktabs}
\usepackage{orcidlink}
\usepackage{xcolor}
\usepackage{listings}

\begin{document}
\title{Fast Radio Bursts and Artificial Neural Networks: a cosmological-model-independent estimation of the Hubble Constant}

\author{Jéferson A. S. Fortunato \footnote{Corresponding author} \orcidlink{0000-0001-7983-1891} }
\email[]{jeferson.fortunato@edu.ufes.br}
 \affiliation{PPGCosmo, CCE, Universidade Federal do Esp\'{\i}rito Santo (UFES), Av. Fernando Ferrari, 540, CEP 29.075-910, Vit\'oria, ES, Brazil}
 \affiliation{N\'ucleo Cosmo-UFES, CCE, Universidade Federal do Esp\'{\i}rito Santo (UFES), Av. Fernando Ferrari, 540, CEP 29.075-910, Vit\'oria, ES, Brazil}
 \affiliation{Institute of Cosmology \& Gravitation, University of Portsmouth, Dennis Sciama Building,
Burnaby Road, Portsmouth, PO1 3FX, United Kingdom}

\author{David J. Bacon \orcidlink{0000-0002-2562-8537}}
\email[]{david.bacon@port.ac.uk}
 \affiliation{Institute of Cosmology \& Gravitation, University of Portsmouth, Dennis Sciama Building,
Burnaby Road, Portsmouth, PO1 3FX, United Kingdom}

\author{Wiliam S. Hip\'olito-Ricaldi \orcidlink{0000-0002-1748-553X}
}
\email[]{wiliam.ricaldi@ufes.br}
\affiliation{N\'ucleo Cosmo-UFES, CCE, Universidade Federal do Esp\'{\i}rito Santo (UFES), Av. Fernando Ferrari, 540, CEP 29.075-910, Vit\'oria, ES, Brazil}
\affiliation{Departamento de Ci\^encias Naturais, CEUNES, Universidade Federal do Esp\'{\i}rito Santo (UFES), Rodovia BR 101 Norte, km. 60,\\
CEP 29.932-540, S\~ao Mateus, ES, Brazil}

\author{David Wands \orcidlink{0000-0001-9509-8386}}
\email[]{david.wands@port.ac.uk}
\affiliation{Institute of Cosmology \& Gravitation, University of Portsmouth, Dennis Sciama Building,
Burnaby Road, Portsmouth, PO1 3FX, United Kingdom}

\begin{abstract}\noindent
Fast Radio Bursts (FRBs) have emerged as powerful cosmological probes in recent years offering valuable insights into cosmic expansion. These predominantly extragalactic transients encode information on the expansion of the Universe through their dispersion measure,  reflecting interactions with the intervening medium along the line of sight. In this study, we introduce a novel method for reconstructing the late-time cosmic expansion rate and estimating the Hubble constant, solely derived from FRBs measurements coupled with their redshift information while employing Artificial Neural Networks. Our approach yields a Hubble constant estimate of $H_0 = 68.2 \pm 7.1\rm \ km \ s^{-1} \ Mpc^{-1}$. With a dataset comprising 21 localised data points, we demonstrate a precision of  $\sim10\%$. However, our forecasts using simulated datasets indicate that in the future it could be possible to achieve precision comparable to the SH0ES
collaboration or the Planck satellite. Our findings underscore the potential of FRBs as alternative, independent tools for probing cosmic dynamics.
\end{abstract}

\maketitle

\section{Introduction}
In the context of the concordance model of cosmology, named $\Lambda$ Cold Dark Matter $(\Lambda \rm CDM)$, the \textit{Hubble tension} emerges as a discrepancy between the estimated value for the present expansion rate of the Universe, characterised by the Hubble constant ($H_0$) when considering early-time versus late-time cosmological probes. Among these probes, two particular methodologies stand out: analysis of the cosmic microwave background (CMB) anisotropies and direct inference from Cepheid stars and type Ia supernovae in the local universe. The Planck Collaboration \cite{aghanim2020planck} meticulously analysed the CMB anisotropies, deriving a value of $H_0=67.4\pm0.5\rm \ km \ s^{-1} \ Mpc^{-1}$ within a $\Lambda \rm CDM$ cosmology. Conversely, leveraging the unique properties of Cepheid stars and type Ia supernovae, \cite{riess2022comprehensive} directly measure $H_0=73.04\pm1.04\rm \ km \ s^{-1} \ Mpc^{-1}$ — a discrepancy of $5\sigma$ with respect to the Planck value. These disparate estimates underscore the urgency of reconciling observations and/or exploring alternative methodologies to achieve a robust determination of the cosmic expansion. Among the various measurements reported in the literature, obtained by analysing different observational probes, the values of $H_0$ usually fall within or close to the above range. For instance, Alam et al~\cite{alam2021completed} using the Baryon Acoustic Oscillation (BAO) scale calibrated with Big Bang Nucleosynthesis measurements, reported a value close to the CMB estimate, $H_0=68.20\pm0.81\rm \ km \ s^{-1} \ Mpc^{-1}$, while Camarena and Marra \cite{camarena2020local} reported $H_0=75.35\pm1.68\rm \ km \ s^{-1} \ Mpc^{-1}$ using the distance-ladder approach. This disparity also appears in other cosmological probes, suggesting either that the considered underlying cosmological model ($\rm \Lambda CDM$) does not provide a consistent description of the Universe, or that there are systematic errors affecting the observational data. 

Over the past decade, researchers have diligently investigated the Hubble tension, which emerged when the first results of the Planck Collaboration were reported in 2013 \cite{ade2013planck, di2021realm}. To this extent, model-dependent and model-independent approaches have been employed for measuring $H_0$, offering complementary insights into the underlying cosmological dynamics. When considering model-dependent techniques, it is necessary to specify a cosmological model to compute distances. These distances are crucial benchmarks against which observational data from cosmological probes are then compared. Conversely, model-independent approaches offer a data-driven way of estimating the underlying theory that governs the observable phenomena. Numerous studies in the literature focus on reconstructing the late-time cosmic expansion history without relying on a specific assumed cosmological model. For instance, Bengaly et al~\cite{bengaly2023measuring} utilised different machine-learning techniques to estimate the Hubble constant, based on direct measurements of the Hubble parameter $H(z)$ inferred from Cosmic Chronometers (CC). This was also done by Seikel et al~\cite{seikel2012reconstruction} using Type Ia Supernovae (SNIa) data and Gaussian Processes (GP) instead. Additionally, recent advancements in Artificial Neural Networks (ANNs) -- another machine learning technique -- have revolutionised cosmological research, offering powerful tools for data-driven analysis, model-independent fitting, and parameter estimation. Unlike GPs, which rely on kernel functions to define correlations between data points, ANNs learn representations and patterns directly from the data, avoiding the issue of overfitting caused by overly strong correlations imposed by kernel choices. This was pointed out by \cite{o2021elucidating} when studying how different kernels lead to tighter constraints on $H_0$, which may be incompatible with the parameter space allowed by certain cosmological models. ANNs hold the potential to address complex cosmological challenges, including estimating $H_0$ and the interpretation of observational data, within the $\Lambda \rm CDM$ framework and beyond. A variety of applications of ANNs have recently been explored in the reconstruction of $H_0$. For instance, Wang et al~\cite{wang2020reconstructing} developed an ANN algorithm and applied it to Cosmic Chronometers data, resulting in a value of $H_0=67.33\pm15.74\rm \ km \ s^{-1} \ Mpc^{-1}$. Similarly, Liu et al~\cite{liu2023measurements}, through the integration of datasets from SNIa and ANNs applied to reconstruct the angular diameter distance from radio quasar data, reported $H_0=73.51\pm0.67\rm \ km \ s^{-1} \ Mpc^{-1}$.

Since their discovery in 2007 by Lorimer et al~\cite{lorimer2007bright}, Fast Radio Bursts (FRBs) have shown promise in providing an alternative cosmological probe, given that they can be used to infer the cosmic expansion and estimate the Hubble constant \citep{macquart2020census, wu20228, zhao2022first, hagstotz2022new}, probe the fraction of baryons in the intergalactic medium (IGM), estimate the cosmic proper distance \cite{yu2017measuring} and other applications. These transient events are characterised by extremely bright and short-duration pulses that occur in the radio spectrum and last for only a few milliseconds. The large dispersion measures ($\rm DMs$) observed in these events, which result from the electron column density along the line of sight, strongly suggest an extragalactic origin. This inference is supported by the observation that the measured $\rm DMs$ far surpass the expected contribution from the Milky Way \cite{lorimer2007bright, petroff2019fast}. To date, hundreds of FRBs have been cataloged, with some exhibiting repeating bursts \cite{zhou2022fast}. Of these, 24 have been precisely localised allowing their host galaxy and redshift to be determined. While some FRB events might have a link to magnetars \citep{bochenek2020fast}, given the possible association between the event FRB200428 and the galactic magnetar SGR 1935+2154, numerous progenitor models have been reported in the literature so far \cite{zhang2020physical, bhandari2020host}. The cosmological origin of these FRBs has made them a prominent observable in the study of cosmology. In a recent study, Liu et al~\cite{liu2023cosmological} combined datasets from FRBs and CCs to estimate the Hubble constant as $H_0 = 71\pm3\rm \ km \ s^{-1} \ Mpc^{-1}$ using a model-independent approach. 

In this paper, we introduce a novel model-independent method for reconstructing the late-time expansion history of the Universe through the Hubble parameter, $H(z)$, estimating $H_0$ via Artificial Neural Networks solely based on the dispersion measures of FRBs. Also, we provide a forecast of the number of FRBs needed to achieve similar precision to that reported by the SH0ES collaboration. The structure of this paper is as follows: Section~\ref{frbs} provides an overview of the fundamental characteristics of Fast Radio Bursts. In Section~\ref{metho} we outline our methodology for reconstructing $H(z)$ and estimating $H_0$. Also, we provide a brief review of Artificial Neural Networks and their implementation for the case considered in this study. Section \ref{results} presents our findings, consisting of both real data analysis and simulation results with forecast data. Finally, Section~\ref{disc} offers a discussion of our results and draws conclusions from our study.

\section{Basic properties of Fast Radio Bursts}\label{frbs}
As a Fast Radio Burst pulse travels toward Earth, it interacts with the intergalactic medium, causing dispersion and a time delay in the arrival of different frequencies that make up the observed signal. This delay is quantified by the Dispersion Measure ($\mathrm{DM}$):
\begin{equation}
\Delta t \propto \left(\nu_{\mathrm{lo}}^{-2} - \nu_{\mathrm{hi}}^{-2}\right) \mathrm{DM},
\end{equation}

\noindent where $\nu_{\mathrm{lo}}$ and $\nu_{\mathrm{hi}}$ represent the lower and higher frequencies of the emitted signal. The $\mathrm{DM}$ parameter is related to the column density of free electrons along the line of sight $l$ to the FRB and is expressed as:

\begin{equation}
\mathrm{DM} = \int \frac{n_e}{(1+z)} dl.
\end{equation}
The observed dispersion measure ($\rm DM_{obs}$) is often estimated by considering the contribution from the local $\rm DM_{\rm local}$ and the extragalactic $\rm DM_{\rm EG}$ environments that the FRB signal crosses:

\begin{equation}\label{dmobs}
    \rm DM_{obs} = \mathrm{DM}_\mathrm{local} + \mathrm{DM}_\mathrm{EG}(z),
\end{equation}
\noindent being 
\begin{equation}
    \mathrm{DM}_{\mathrm{local}}=\mathrm{DM}_{\mathrm{ISM}}+\mathrm{DM}_{\mathrm{halo}},
\end{equation}

\noindent and
\begin{equation}
    \mathrm{DM}_{\mathrm{EG}} = \mathrm{DM}_{\mathrm{IGM}}+\frac{\mathrm{DM_{\mathrm{host}}}}{(1+z)},
\end{equation}

\noindent where $\rm DM_{\rm ISM}$ is the interstellar medium contribution and $\rm DM_{\rm halo}$ is the Milky Way surrounding halo component. The former, $\rm DM_{\rm ISM}$, has been extensively studied and commonly calculated using galactic electron distribution models such as NE2001 \cite{cordes2002ne2001} and YMW16 \cite{yao2017new} and then subtracted from the observed dispersion measure. In this study, we adopted the NE2001 model. Recent studies have indicated that the YMW16 model may overestimate $\rm DM_{\rm ISM}$ at low Galactic latitudes \cite{ocker2021constraining}. On the other hand, $\rm DM_{\rm halo}$ is the Milky Way galactic halo contribution that is estimated in the range $50<\rm DM_{\rm halo}< 80 ~\rm pc~cm^{-3}$ from the Sun to a distance of 200 kpc \cite{prochaska2019probing}. We employ $\rm DM_{\rm halo}=50 ~\rm pc~cm^{-3}$ in our calculations as a conservative choice. $\rm DM_{\rm IGM}$ is the intergalactic medium contribution, which is the main contribution to the observed dispersion measure and contains the cosmological dependence.  Previous works \cite{mcquinn2013locating} have shown that this contribution accounts for a scatter around the mean $\rm \langle DM_{\rm IGM} \rangle$, $100 - 400 ~\rm pc~cm^{-3}$  at z = 0.5 - 1, and $\rm DM_{\rm IGM}$ is in general inhomogeneous due to the non-smooth electron distribution along the signal path.  The mean is calculated through the so-called Macquart Relation \cite{macquart2020census} and depends upon the Hubble rate:
\begin{equation}\label{DM_igm}
{\rm \langle DM_{IGM} \rangle } = \left(\frac{3c}{8\pi Gm_{\rm p}}\right){\rm \Omega_b} H_0\int^z_0\frac{(1+z)f_{\rm IGM}(z)f_{\rm e}(z)}{E(z^\prime)}dz^\prime,
\end{equation}
\noindent 
where $E(z)=H(z)/H_{0}$. We denote the cosmic baryon density by $\rm \Omega_b$,  $m_{\rm p}$ is the proton mass, and the baryon mass fraction in the $\rm IGM$ is $f_{\rm IGM}$. 

The electron fraction $f_{\rm e}(z) = Y_{\rm
H}X_{\rm e,H}(z)+\frac{1}{2}Y_{\rm He}X_{\rm e,He}(z)$, where we consider an $\rm IGM$ with a hydrogen mass fraction $Y_{\rm
H}=0.75$ and a helium mass fraction $Y_{\rm He}=0.25$. For simplicity, we assume that both hydrogen and helium are fully ionized at $z<3$. This assumption is widely used in FRB cosmology to estimate the IGM component. While the IGM may not be fully ionized, especially for helium at the redshifts of the FRB sample considered in this study, it is reasonable to assume that helium is fully ionized without significant loss of accuracy. Even if helium were not fully ionized, its impact on the results would likely be minimal, as the majority of the DM contribution comes from fully ionized hydrogen. Then, we set the ionisation fractions for each species as $X_{\rm e,H}=X_{\rm e,He}=1$. We shall return to $f_{\rm IGM}$ later. 

$\rm DM_{\rm host}$  is related to the host galaxy environment and it has to be scaled with redshift to account for the cosmic expansion. Although some works treat this component as a constant value, see \cite{tendulkar2017host} for example, where the authors set  $\rm DM_{\rm host}=100~  \rm pc ~cm^{-3}$, this can lead to an overestimation of the dispersion measure at lower redshifts and result in negative values for the IGM component, which has no physical meaning. To avoid this issue, we adopt a parametrization introduced in  \cite{zhang2020dispersion}, based on the IllustrisTNG cosmological simulation,

\begin{equation}\label{dmh}
    \rm DM_{\rm host} = \rm A\left( 1+ z \right)^\alpha,
\end{equation}

\noindent where A and $\alpha$ are free parameters to be fitted based on the host local environment properties, which shows that for repeating FRBs, depending on the host galaxy type (spiral or dwarf galaxy), the mean $\rm DM_{\rm host}$ can be $34.72 (1+z)^{1.08}~  \rm pc ~cm^{-3}$ or $96.22 (1+z)^{0.83}~  \rm pc ~cm^{-3}$, and for non-repeating FRBs, $32.97 (1+z)^{0.84}~  \rm pc ~cm^{-3}$.

\section{Method}\label{metho}
In this section, we present the method used to reconstruct the late-time expansion history through the Hubble parameter $H(z)$ and estimate the Hubble constant $H_0$ by using only localised Fast Radio Bursts data and Artificial Neural Networks. We start by presenting some of the equations relevant to our approach.
\subsection{Estimating the Hubble constant $H_0$}
Dispersion measures originating from the intergalactic medium and their redshift dependence are related to the Hubble parameter. Therefore Eq.~(\ref{DM_igm}) can help us to reconstruct the late-time cosmic expansion history, $H(z)$, by considering the derivative of ${\rm \langle DM_{IGM} \rangle}$  with respect to redshift: 
\begin{equation}\label{M!}
    H(z)=\left(\frac{3c}{8\pi G m_p}\right)10^4 \Omega_b h^2 \left(1+z\right) f_{\rm IGM}(z) f_e(z) \left(\frac{d\langle \rm DM_{\rm IGM}\rangle}{dz}\right)^{-1}.
\end{equation}

Thus, the key to reconstruct $H(z)$ from the localised FRB datasets is to know $\left(\frac{d\langle \rm DM_{\rm IGM}\rangle}{dz}\right)^{-1}$. Some approaches to compute $\langle \rm DM_{\rm IGM}\rangle$ and its derivative with respect to redshift directly from the data have been explored in the literature. For example, one approach uses a dataset split into bins to compute the mean of the redshifts and dispersion measures in each bin \cite{wu2020new}. On the other hand, although \cite{liu2023cosmological} also uses
the division into bins, they incorporate the Markov Chain Monte Carlo (MCMC) method with a combination of a continuous piece-wise linear function to approximate $\langle \rm DM_{\rm IGM}\rangle$. In this work however we use a completely different approach considering ANNs,  which allow us to construct $\left(\frac{d\langle \rm DM_{\rm IGM}\rangle}{dz}\right)^{-1}$ with no a priori assumptions and in a completely data-driven way.

We start by reconstructing the average value for the IGM component $\langle \rm DM_{IGM} \rangle$ via ANNs as described in the next subsection. The inputs are the redshift and the $\rm DM_{\rm IGM}$ estimated from the observed dispersion measure, subtracting the other contributions discussed in the previous section:

\begin{equation}\label{dmigmest}
\mathrm{DM}_{\rm IGM} = \rm DM_{\rm obs} - \rm DM_{\rm local} - \rm DM_{\rm host}(1+z)^{-1},
\end{equation}
and its associated uncertainty given by
\begin{equation}\label{error}
\sigma_{\rm IGM}(z) = \sqrt{\sigma^2_{\rm obs}(z) + \sigma^2_{\rm local} + \sigma^2_{\rm\Delta IGM}(z) + \left(\frac{\sigma_{\rm host}(z)}{1+z}\right)^2},
\end{equation}
where $\sigma_{\rm obs}$ and $\sigma_{\rm host}$ represent the errors related to $\rm DM_{\rm obs}$ and $\rm DM_{\rm host}$, respectively, while $\sigma_{\rm local}$ is the sum of $\rm DM_{\rm ISM}$ and $\rm DM_{\rm halo}$ errors. This equation provides a framework for characterising the dispersion in $\langle \rm DM_{IGM} \rangle$ across different sight-lines. For our calculations, we follow the approach outlined by \cite{hagstotz2022new}, where $\sigma_{\rm local}\approx 30~\rm{pc ~cm}^{-3}$ and the uncertainty of the host component $\sigma_{\rm host}$ is computed using the upper and lower limits of two fitting parameters of Eq.~(\ref{dmh}): A and $\alpha$. Due to the inherent inhomogeneity of baryonic matter in the IGM, a large dispersion is observed around the mean value $\langle \rm DM_{IGM} \rangle$. For instance, McQuinn et al. \cite{mcquinn2013locating} reported a scatter ranging from $100-400~\rm pc~cm^{-3}$ within the redshift range $\rm z = 0.5 - 1.0$. To address this variability, we adopt a power-law function derived from the findings of \cite{mcquinn2013locating}, as elaborated in \cite{qiang2020reconstructing}:

\begin{equation}\label{scatterigm}
\sigma_{\rm\Delta IGM}(z)=173.8~ z^{0.4}~\rm pc~cm^{-3}.
\end{equation}

Once the average $\langle \rm DM_{IGM} \rangle$  is reconstructed from the data, the derivative $\frac{d\langle \rm DM_{\rm IGM}\rangle}{dz}$ and its inverse are computed to reconstruct $H(z)$ using Eq.~(\ref{M!}). Finally, to estimate the Hubble constant, the extrapolated value of $H(z)$ at $z = 0$ is taken. The uncertainty is propagated according to the following equation:
\begin{equation} \label{errorhz}
    \sigma_{H} = H(z)~  \sqrt{ \left(\frac{\sigma_{\Omega_b h^2}}{\Omega_b h^2}\right)^2 + \left(\frac{\sigma_{f_{\rm{\rm IGM}}(z)}}{f_{\rm{IGM}}(z)}\right)^2 +   \left(\frac{\sigma_{\frac{d z}{\langle \rm DM_{IGM} \rangle}}}{\frac{d z}{\langle \rm DM_{IGM} \rangle}}\right)^2 }\,,
\end{equation}
\noindent which also is extrapolated to find the error $\sigma_{H_0}$ for $H_0$ estimation. We emphasise that our approach relies solely on the localised FRBs data.

\subsection{Artificial Neural Networks}\label{anns}

\begin{figure}[htbp]
\centering
\includegraphics[width=.4\textwidth]{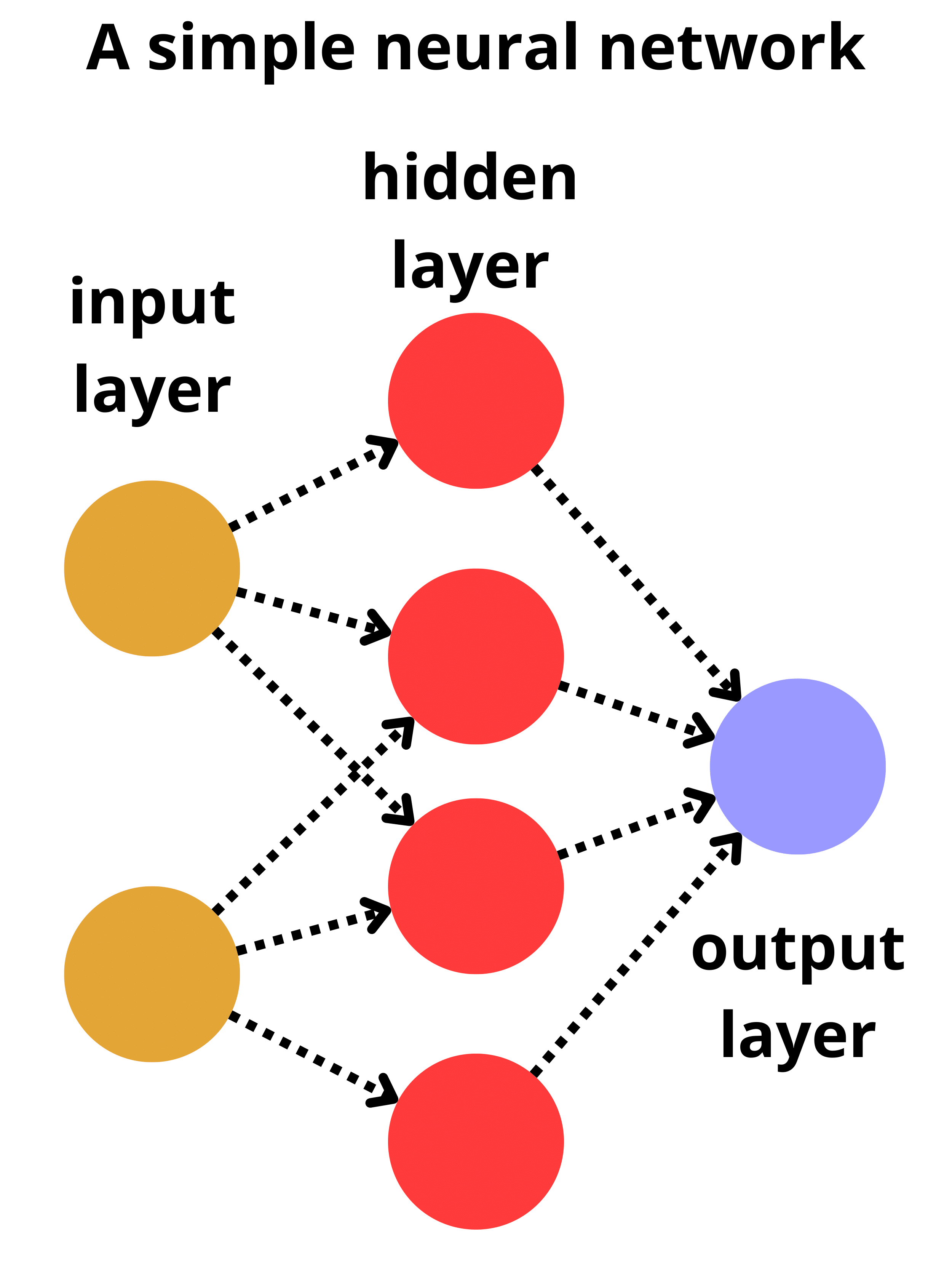}
\caption{Schematic representation of a simple Artificial Neural Network architecture, depicting the input layer, hidden layer, and output layer. The structure of the network displays the flow of information through interconnected nodes. \label{figi}}
\end{figure}

Artificial Neural Networks are computational machine learning tools that simulate the learning mechanisms observed in biological systems. The computational units that make up an ANN, used for processing input information, are called neurons \cite{aggarwal2018neural}. These artificial neurons work by taking quantities available in their inputs, processing them according to their operational function, and producing a response as an output considering their activation function \cite{da2017artificial, priddy2005artificial}. The most common neuronal structure of an ANN consists of an input layer connected to a hidden layer (containing a certain number of neurons), or a group of sequential hidden layers, and an output layer \cite{dialektopoulos2022neural}. See Figure~\ref{figi} for the scheme used in this work. The connections between neurons are related to quantities referred to as weights, implemented by vectors and matrices. Each input variable is scaled with a weight in order to quantify its relevance with respect to the function calculated at that neuron \cite{da2017artificial}. The ANN calculates a function of the inputs by propagating computed values from the input neurons to output neuron(s), with the weights as intermediary parameters. The learning process occurs by adjusting these weights to refine the computed function for improved predictions in subsequent iterations \cite{aggarwal2018neural}.

To be more precise, following the methodology outlined by \cite{liu2022machine}, each layer receives a vector from the previous layer as an input. Then a linear transformation is applied followed by the application of an activation function, which calibrates the output within a range of reasonable values based on its own functional form \cite{da2017artificial}, then propagates the current result to the next layer. In the machine learning framework, the basic idea is that the following relation gives the prediction function $z$:
\begin{eqnarray}\nonumber
z\left( \mathbf{w; x}\right)&=&\mathbf{x}\mathbf{w}+b;\\\nonumber
    &=&w_1x_1+\cdot\cdot\cdot+w_px_p+b\\\nonumber
    &=&\sum^p_iw_ix_i+b;\\\nonumber
    y &=& f(z_{i+1}),
\end{eqnarray}
\noindent with $z\left( \mathbf{w; x}\right)$ read as a function of $\mathbf{w}$ for a given $\mathbf{x}$. The vectors are defined as follows: 

\begin{eqnarray}\nonumber
    \mathbf{x}&=&\left[ x_1, x_2, ..., x_p \right];\\\nonumber
    \mathbf{w}&=&\left[ w_1, w_2,..., w_p \right]^\top,
\end{eqnarray}

\noindent where $x_i(i=1,2...,p)$ are the feature variables used as inputs of the i-th layer. On the other hand, $\mathbf{w}$ and $b$ represent the weights and the bias, respectively to be optimised. These two values can provide information about the performance of the predictions of the ANN \cite{gomez2021cosmological} when comparing the predicted values to the actual values of the considered dataset. The quantity $z_{i+1}$ is the intermediate vector after applying the linear transformation, and $f$ is the activation function. In this work, we choose the Rectified Linear Unit (ReLU) as the activation function to limit the neuron output:

\begin{align}\label{relu}\nonumber
&f(z) = 
\begin{cases}
0 & z < 0 \\
z & z \geq 0
\end{cases} \quad ,
\end{align}

\noindent which is widely adopted in machine learning applications due to its ability to address the gradient vanishing problem commonly encountered in deep neural networks, promoting more stable and efficient training processes and enabling the network to learn complex relationships within the data more effectively.

The training process is a fundamental step to determine if the ANN performs well and generalizes well to unseen data, that is, the ability to predict the underlying patterns based on new data. For this reason, it is necessary to randomly split the considered dataset into two different groups: the training and the test sets. The training set is required to train and optimize the machine learning model by updating the weights and bias every interaction aiming to minimize the \textit{loss function}, which quantifies the difference between the predicted values, derived from the current weights and biases, and the actual data points in the training set. Through techniques such as backpropagation, the model is progressively adjusted to reduce this discrepancy, then enhancing its accuracy on the training data. The test set is kept for the final evaluation of the performance of the model; this way it is possible to determine its generalization ability and performance on new, previously unseen data points \cite{bishop2006pattern, liu2022machine, raschka2020machine}.

\subsection{ANN implementation}\label{implementation}

Throughout this investigation, we utilised an ANN architecture known as Multilayer Perceptron (MLP). The MLP model is implemented in Python language by the scikit-learn library, which is a powerful tool designed to handle various machine learning applications \cite{scikit-learn}. With this algorithm and using the measured redshifts and $\rm DM_{IGM}$ as inputs, we are able to reconstruct the mean relation $\langle\rm DM_{IGM}\rangle$ as a function of $z$. To ensure robust model performance, we first partitioned our dataset into training and testing sets using a standard 75-25 split ratio -- although we also explored alternative configurations such as 50-50, 60-40, and 90-10, yielding results comparable to those detailed in Section \ref{results}. The standard split led to superior performance, given that for a better training process it is necessary to include more data points. Also, we fixed a random state of 152. This parameter is used to control the randomness of machine learning algorithms, such as train-test splitting or shuffling of data in cross-validation. We are able to obtain reproducible results by fixing the seed of the random number generator. Subsequently, we conducted hyperparameter tuning for the MLPRegressor model using the module GridSearchCV. To identify the optimal configuration, we employed different combinations of various parameters of the module, including activation functions, solvers, and learning rates. Specifically, we varied the sizes of the hidden layers from 10 to 240 neurons in increments of 10. Through this systematic search process, we aimed to find the best-performing model configuration for our particular FRB dataset. Finally, we fit the MLPRegressor model to the training data, ensuring that it learned the underlying patterns in our dataset to make accurate predictions. The algorithm snippet used to perform our method is given by:

\begin{center}
\begin{lstlisting}
# Splitting the dataset into training and testing sets
z_train, z_test, dm_train, dm_test = train_test_split(z, dm, test_size=0.25, random_state=152)

# Hyperparameter tuning using GridSearchCV for MLPRegressor
regr = GridSearchCV(MLPRegressor(activation='relu', solver='lbfgs', learning_rate='adaptive', max_iter=200, random_state=152),
                   param_grid={'hidden_layer_sizes': np.arange(10, 250, 10)},
                   cv=2, refit=True) 

# Fitting the model to the training data
regr.fit(z_train, hz_train.reshape((len(hz_train),)))

\end{lstlisting}
\end{center}

Although this algorithm is suitable for reconstructing underlying functions based on the input data, it lacks a built-in mechanism to quantify estimation uncertainties. In light of this, we adopt bootstrap resampling of the dataset as the methodology to enable us to estimate the inherent uncertainty associated with the reconstructed $\langle \rm DM_{IGM} \rangle$. Such a method has already been implemented in other works in a similar fashion \cite{bengaly2023measuring, gong2024ksz}. We employ then $100$ resamples to reconstruct the statistical confidence regions. This strategy not only reduces computational complexity but also optimizes the use of available datasets if it is comprised of a small sample, enabling a more straightforward estimation of uncertainties.


\section{Data and results}\label{results}
Although numerous Fast Radio Bursts have been documented, so far only 24 have had their redshifts determined. Here we use a subset of 21 from the available FRBs measured together with their host galaxy redshifts, compiled by \cite{yang2022finding} and listed in Table~\ref{dataset}. We exclude FRB200110E because it carries little cosmological information due to the fact that it is the closest extragalactic FRB detected to date, only $3.6~\rm M pc$ distant, located in a globular cluster in the M81 galaxy \cite{kirsten2022repeating, bhardwaj2021nearby}. Also, we removed FRB210117 from our analysis due to its observed dispersion measure significantly exceeding the expected value for its redshift, classifying it as an outlier. This can be caused by conditions within its host environment \cite{bhandari2023nonrepeating}. Following the methodology described in section \ref{metho}, the estimated value of $\rm DM_{\rm IGM}$ for FRB181030, at redshift $z = 0.0039$, became negative, which is non-physical. Therefore, we excluded this event from our study.

\begin{figure}[htbp]
\centering
\includegraphics[width=0.6\textwidth]{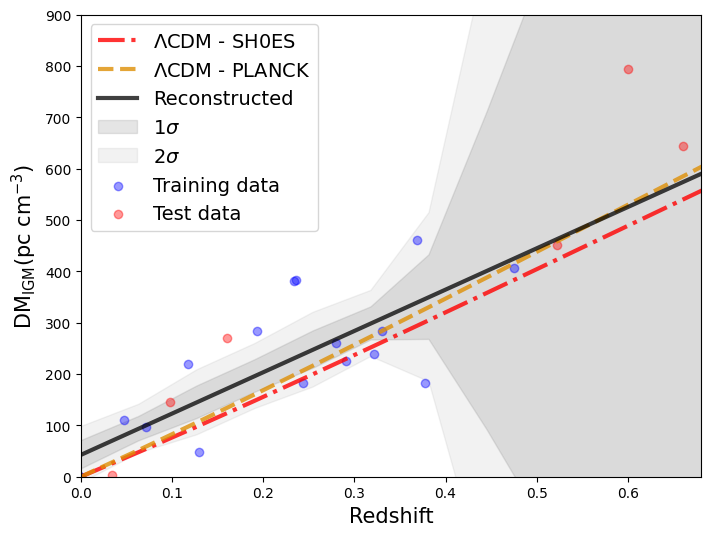}
\caption{Reconstructed mean intergalactic component of the dispersion measure $\langle\rm DM_{IGM}\rangle$ against redshift employing Artificial Neural Networks, depicted with corresponding statistical confidence levels. We plot also the training and testing datasets used for network training. For comparison theoretical curves for $\langle\rm DM_{IGM}\rangle$ are shown using the $H_0$ values from the Planck collaboration and the SH0ES team in a $\Lambda$CDM cosmology with $\Omega_m=0.315$~\cite{aghanim2020planck}.}\label{dmrec}
\end{figure}

$\mathrm{DM}_{\rm IGM}$ and its associated error are computed using the data from Table~\ref{dataset} for each redshift using Eqs.~(\ref{dmigmest}) and~(\ref{error}). The ANN process, with the optimized hyperparameters as discussed in subsection~\ref{implementation}, is then applied to reconstruct the average $\langle \rm DM_{IGM} \rangle$, which is presented in Figure~\ref{dmrec} (black continuous line) and its 1$\sigma$ and 2$\sigma$ regions (gray shaded regions). For the sake of comparison, Figure~\ref{dmrec} also shows theoretical curves for $\Lambda$CDM with the best-fit parameters found by the SH0ES collaboration~\cite{riess2022comprehensive} and by the latest  Planck satellite release~\cite{aghanim2020planck}, alongside the training and testing sets. Notice that the $\Lambda$CDM model with both $H_0$ values (SH0ES and Planck) is consistent with the reconstructed $\langle \rm DM_{IGM} \rangle$ within a 1$\sigma$ confidence level.

\begin{table}
	\small
	\begin{center}
            
		\begin{tabular}{ccccc}
                
			\hline
			Name & Redshift & DM$_{\rm obs}$   & Reference \\
			&  & $(\rm pc \ cm^{-3})$  & &  \\
			\hline
			FRB 121102 &    0.19273 &$557$	    & \cite{spitler2016repeating}\\
			FRB 180301 &    0.3304  &$534$           & \cite{bhandari2022characterizing}\\
			FRB 180916 &	0.0337  &$348.8$	      & \cite{marcote2020repeating}\\
			FRB 180924 &    0.3214  &$361.42$        & \cite{bannister2019single}\\
			FRB 181030 &    0.0039  &$103.5$      & \cite{bhardwaj2021local}\\
			FRB 181112 &    0.4755	&$ 589.27$      & \cite{prochaska2019low}\\
			FRB 190102 &    0.291 	&$ 363.6$	       & \cite{bhandari2020host}	\\
			FRB 190523 &    0.66    &$ 760.8$          & \cite{ravi2019fast} \\
			FRB 190608 &    0.1178  &$ 338.7$	       & \cite{chittidi2021dissecting}	\\
			FRB 190611 & 	0.378   &$ 321.4$	     & \cite{day2020high}  \\
			FRB 190614 &    0.6     &$959.2$            & \cite{law2020distant}\\
			FRB 190711 &    0.522 	&$ 593.1$     & \cite{heintz2020host}\\
			FRB 190714 &    0.2365	&$ 504.13$         & \cite{heintz2020host}\\
			FRB 191001 &	0.234   &$ 506.9$        & \cite{heintz2020host}\\
			FRB 191228 &    0.2432  &$ 297.5$         & \cite{bhandari2022characterizing}\\
			FRB 200430 &    0.16	&$ 380.25$      & \cite{heintz2020host}\\
			FRB 200906 &    0.3688  &$ 577.8$          & \cite{bhandari2022characterizing}\\
			FRB 201124 & 	0.098   &$ 413.52$      & \cite{fong2021chronicling}\\
            FRB 210117 & 0.2145 & $730.0$& \cite{james2022measurement} \\
            FRB 210320 & 0.2797 & $384.8$& \cite{james2022measurement} \\
		FRB 210807 & 0.12927 & $251.9$ & \cite{james2022measurement}\\
		FRB 211127 & 0.0469 & $234.83$ & \cite{james2022measurement} \\
		FRB 211212 & 0.0715 & $206.0$ & \cite{james2022measurement} \\
			\hline
		\end{tabular}
		\vspace{0.5cm}
  \caption{ Properties of localised FRBs.} \label{dataset}
	\end{center}
\end{table}

\begin{figure}[htbp]
\centering
\includegraphics[width=0.6\textwidth]{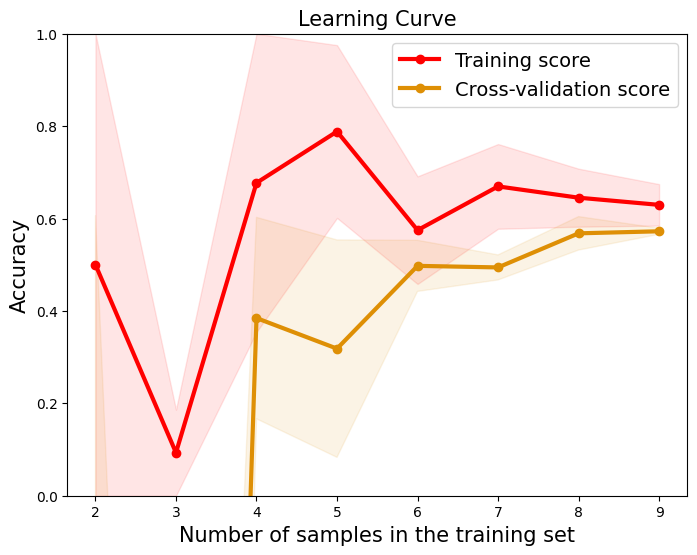}
\caption{Learning curve for the median values of the scores computed for the training set.}\label{learninc}
\end{figure}

As we are interested in a completely data-driven process to measure $H_0$, we have used the real data for both the training and validation processes. Given that the dataset is small, we need to assess the performance of the learning process. This is achieved by setting the cross-validation parameter to $\text{cv} = 2$, which splits the available training data into $(S-1)/S$ portions, where $S$ is the number of folds, for training and keeps the remaining portion for validation \cite{bishop2006pattern}. Then, we used the \texttt{learning\_curve} function from \texttt{scikit-learn} to plot the learning curves, which allows us to vary the number of data points within the training set and compute scores for each fold. In this case, we use accuracy as the scoring metric. The \textit{training score} and the \textit{cross-validation score} are plotted in Figure~\ref{learninc}.

The \textit{training score}, shown by the red line, indicates how well the model fits the training data. Initially, the training score starts at 0.5 with the smallest training set size and then dips to a lower value, before rapidly increasing and fluctuating. The rapid fluctuations and high values suggest that the model is overfitting for smaller training sets. However, as the number of training samples increases, the training score stabilises around 0.6, indicating that the model fits the training data moderately well.

On the other hand, the \textit{cross-validation score}, represented by the yellow line, measures how well the model generalises to unseen data. The cross-validation score starts low, around -1, but gradually increases with the training set size. The fluctuations reflect instability when the model has fewer training samples from which to learn. As the training set size grows, the cross-validation score improves, suggesting that the model generalises better to unseen data when more training samples are available. However, the gap between the training and cross-validation scores, although reduced, remains, indicating that there is still room for better generalisation as more data is incorporated. The idea is, that as the training set size increases, the large gap between the training score and the cross-validation score should narrow to reflect better model generalisation.

\begin{figure}[htbp]
\centering
\includegraphics[width=0.6\textwidth]{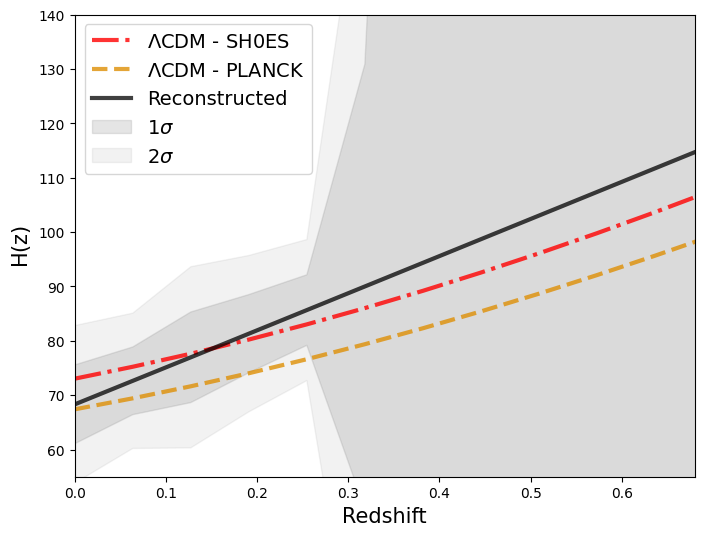}
\caption{The late-time cosmic expansion rate $H(z)$ reconstructed by employing Artificial Neural Networks, displayed with $1~\sigma$ and $2~\sigma$ statistical confidence levels. The figure includes the curves predicted by the $\Lambda\rm CDM$ concordance model, using $H_0$ values from the Planck collaboration or the SH0ES team for comparison, setting $\Omega_m=0.315$~\cite{aghanim2020planck}.}\label{hzrec}
\end{figure}

After we obtain $\rm \langle DM_{IGM} \rangle$, its derivative with respect to redshift is computed to reconstruct  $\left(\frac{d\langle \rm DM_{\rm IGM}\rangle}{dz}\right)^{-1}$. Finally, the Hubble function is obtained using Eq.(\ref{M!}) and its error using Eq.(\ref{errorhz}). During the process, we have used $\Omega_b h^2=0.02235\pm0.00049$. This value comes from primordial Big Bang Nucleosynthesis (BBN)~\cite{cooke2018one}, where $\Omega_b h^2$ can be estimated by observing the primordial deuterium-hydrogen abundance ratio. On the other hand, we use $f_{\rm IGM} = 0.83\pm0.06$ in agreement with the value found by \cite{yang2022finding}, estimated using 22 localised FRBs. The reconstructed $H(z)$ function during the late time universe is presented in Figure~\ref{hzrec} (black solid curve) alongside its 1$\sigma$ and 2$\sigma$ regions in grey. Figure~\ref{hzrec} also shows the curves for both the best-fit set of parameters found by the SH0ES collaboration and by the latest Planck satellite release. We can observe that the $\Lambda$CDM model with both $H_0$ values (SH0ES and Planck) is consistent with the reconstructed $H(z)$ within a 2$\sigma$ confidence level. At 1$\sigma$, the reconstructed curve is quite consistent with the $H_0$ value measured by Planck, though $H(z)$ slightly deviates at higher values of $z$. An extrapolation to $z=0$ of the reconstructed $H(z)$ and its error gives us the value of the constant Hubble, $H_0 = 68.2 \pm 7.1$. The precision of the estimated $H_0$ using only localised FRBs is $\sim 10\%$. This precision could change slightly depending on the different reported values for $f_{\rm IGM}$. For instance, $f_{\rm IGM}=0.82\pm0.07$, as found in \cite{fortunato2023cosmography}, results in a precision of $\sim 11\%$, while $f_{\rm IGM}=0.82\pm0.04$, reported by \cite{wu2020new}, results in $\sim 9\%$ precision.

Following \cite{zhou2014fast, hashimoto2021revealing}, we tested the impact of a linearly increasing intergalactic medium fraction, \( f_{\rm IGM} \), as a function of redshift, using the relation \( f_{\rm IGM} = 0.053z + f_{\rm IGM, 0} \), with \( f_{\rm IGM, 0} = 0.82 \). Our results showed no significant differences when applying this model. Other works \cite{wang20238, lin2023probing} have explored different parameterizations, such as \( f_{\rm IGM} = f_{\rm IGM, 0}\left(1 + \alpha \frac{z}{z+1}\right) \), finding that \( f_{\rm IGM} \) exhibits minimal redshift dependence at low  \( z \). Similarly, varying the value of the parameter \( \Omega_bh^2 \) did not produce noticeable changes in our results, whether we used the value from the latest Planck data release or the one derived from Big Bang nucleosynthesis.

The existing sample of localised FRBs is not sufficient to reach a statistical precision comparable to some of the other probes, for example, SNIa. However, considerable effort is going into the construction of instruments to improve the quantity of data that will be available in the future. This includes two upcoming radio telescopes: the Square Kilometre Array (SKA) \cite{zhang2023cosmology} and the Baryon Acoustic Oscillations from Integrated Neutral Gas Observations (BINGO) project \cite{abdalla2022bingo, santos2023bingo}. The SKA promises vast detection capability for FRBs, and while BINGO focuses on 21-cm HI line detection, it also holds significant potential for FRB research. Therefore, the question arises: what is the impact of the future FRBs datasets, much larger than those currently available? To answer this question we will assess our methodology using Monte Carlo simulated datasets.

\begin{figure}[htbp]
\centering
\includegraphics[width=0.6\textwidth]{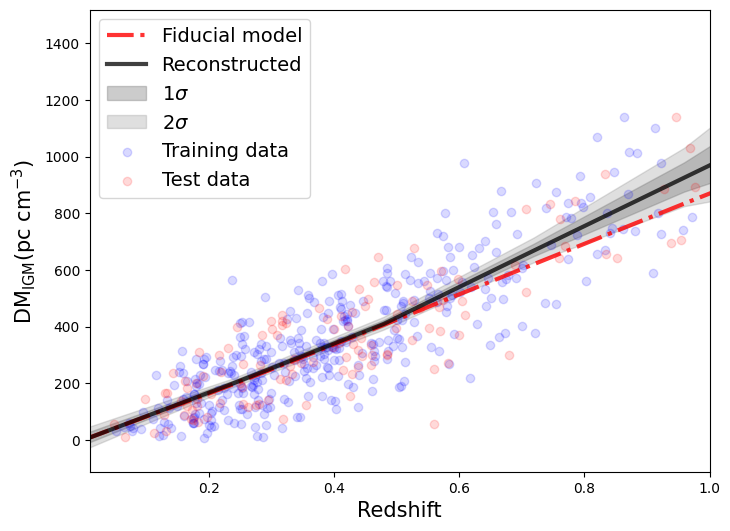}
\caption{Reconstruction of the mean intergalactic component of the dispersion measure $\langle\rm DM_{IGM}\rangle$, shown along with the statistical confidence contours, for one of the simulations from our 100 mock datasets. We show the training and testing datasets. The same curve is plotted for the fiducial model used to simulate the FRBs for comparison.}\label{dmrecsim}
\end{figure}

To generate the mock data it is necessary to choose a fiducial cosmological model. Here we use the flat $\Lambda$CDM model with the cosmological parameters $H_0 = 70$, $\Omega_b=0.049$, $\Omega_m=0.3$ and $\Omega_{\Lambda}=1-\Omega_m$. Then, we follow the methodology outlined in references \cite{yu2017measuring} and \cite{liu2023cosmological}. The FRB dataset has information on $\rm DM_{\rm obs}$ and its correspondent redshift. We assume the redshift distribution can be described by $f(z)\approx z^2 \exp(-\alpha z)$ within the range $0<z<1$, where $\alpha =7$ \cite{hagstotz2022new}.  $\alpha$ is a parameter to control the depth of the mock sample, working as a cutoff placing, with most of the data points in the range z = 0.3 -- 0.5. In the FRB context, it is common to simulate  $\rm DM_{\rm EXT}=\rm DM_{\rm obs}-DM_{\rm loc}=\rm DM_{\rm IGM}+DM_{\rm host}$, employing randomly sampled $\rm DM_{\rm EXT}^{\rm sim}=\rm DM_{\rm IGM}^{\rm sim}+DM_{\rm host}^{\rm sim}$ as the observed quantity.  From Eq.~(\ref{DM_igm}), ${\rm \langle DM_{IGM}^{fid} \rangle }$ is computed using the fiducial model, then we sample $\rm DM_{\rm IGM}^{\rm sim}$ from a normal distribution $\mathcal{N}\left({\rm \langle DM_{IGM}^{fid} \rangle }, \sigma_{\rm DM_{\rm IGM}^{fid}} \right)$, where $\sigma_{\rm DM_{\rm IGM}^{fid}}=\sigma_{\rm\Delta IGM}(z)$, as given by equation (\ref{scatterigm}). The host component $\rm DM_{\rm host}^{\rm sim}$ is drawn from a lognormal distribution:
\begin{equation}\label{Phost}
P({\rm DM_{host}}) = \frac{1}{{\rm DM_{host}} \sigma_{\rm host} \sqrt{2\pi}} {{\exp}}\left(-\frac{({\rm ln} ({\rm DM_{host}}/\mu))^2}{2\sigma_{\rm host}^2}\right)\,.
\end{equation}
Following \cite{macquart2020census} we consider ${\mu}$ in the range $20-200~\rm pc~cm^{-3}$ and $\sigma_{\rm host}$ in the range $0.2-2.0$. Using our methodology applied to 100 simulated datasets each comprised of 500 mock data points (see Figure \ref{dmrecsim} for a visual inspection of one of the reconstructions), we derived a series of estimates for the Hubble constant. Employing the minimum $\chi^2$ method, with flat prior over $H_0$ ranging from $0$ to $100$, we obtained $H_0 = 69.4\pm0.8$. This result reflects a statistical precision of $\sim 1.15\%$, closely aligned with the precision reported by the SH0ES collaboration, which stands at $\sim 1.42\%$. Note that we use a value of $f_{\text{IGM}} = 0.8423 \pm 0.0110$, as estimated by \cite{santos2023bingo}. This estimate was obtained by simulating 500 FRB data points with a redshift distribution similar to that used in this work, ensuring consistency in the simulated data length.

\section{Discussions}\label{disc}
Although some astrophysical characteristics of the FRBs (e.g features of the host, the real impact of the electron distribution inhomogeneity, etc) are not fully understood, in recent years, their potential importance in cosmology has become clear, providing an alternative probe to test fundamental physics, constraining cosmological models and estimating cosmological parameters. Studies probing the underlying physical FRBs mechanisms will improve our understanding of their potential and possible limitations. In this sense model-independent analyses are preferable because they should universally valid. In this work, we reconstruct the Hubble function $H(z)$ and estimate the Hubble constant $H_0$ using ANN techniques. We emphasise that, unlike other model-independent approaches involving FRBs, $H_0$ was estimated only using current observations of localised FRBs without relying on additional probes. This is acheived by subtracting the different contributions to the observed dispersion measure, in such a way as to leave only the $\rm DM_{\rm IGM}$ contribution for each redshift. Our method is based on the known relation between the average of the IGM dispersion measure and the Hubble parameter.
The ANN algorithm is applied after an optimisation process to find the optimal hyperparameters set for our particular dataset.

The value we found for Hubble constant is $H_0 = 68.2 \pm 7.1$, corresponding to $\sim 10\%$ precision, slightly
further from the SH0ES result than the CMB one. Similar conclusions have been reached in previous works \cite{fortunato2023cosmography, wu20228, hagstotz2022new}, however, our findings hold for a minimal number of assumptions, are completely data-driven, and use only localised FRBs. More data is necessary to confirm or discard this result. Nonetheless, the precision we found here is of the order or slightly tighter than reported in some previous works \cite{fong2021chronicling, wu20228}. 

In the FRB context, leading contributions for the uncertainty come from three sources: i) systematics and instrumental effects, ii) a poor understanding of the DM fluctuation due to variation in the electrons' distribution along the signal path and no exact knowledge about host galaxy contribution, and iii) a small number of localised FRBs in the current dataset. Unfortunately to diminish the uncertainty coming from the first two sources we have to wait for future developments in both FRBs physics and observations. If in future we increase the number of localised FRBs, then results seem promising. We show this by simulating $\sim 500$ FRB events with corresponding redshifts and found that they could increase precision to the same order of precision as current SNIa, i.e., $\sim 1\%$.  The projected increase in the number of events appears feasible in coming years, particularly with the work of ongoing projects such as the Canadian Hydrogen Intensity Mapping Experiment (CHIME) in Canada \cite{amiri2021first}, the Five-hundred-meter Aperture Spherical Telescope (FAST) in China \cite{li2018fast} and the More Karoo Array Telescope (MeerKAT) \cite{jonas2016meerkat}. The MeerKAT is a precursor of the already mentioned SKA, consisting of an interferometric array consisting of 64 dishes, each with a diameter of 13.96 meters, situated in the Karoo desert of South Africa. These initiatives, alongside others outlined in earlier sections, are poised to contribute significantly to the growing pool of observed FRB events.

In summary, our results are optimistic and promising, showing the potential of future FRB observations to constrain $H_0$ with high precision, through a completely data-driven approach. This suggests that FRBs could emerge as an alternative, independent method for estimating $H_0$, shedding new light on the challenging problem of the Hubble tension in the future.

\acknowledgments

JASF thanks FAPES and CNPq for their financial support. WSHR thanks FAPES (PRONEM No 503/2020) for the financial support under which this work was carried out. The authors thank Marcelo V. dos Santos and Yang Liu for providing valuable insights into some research questions. Also, JASF is grateful for the hospitality of the Institute of Cosmology and Gravitation of the University of Portsmouth where most of this work was developed.
This work was supported by the Science and Technology Facilities Council (grant number ST/W001225/1).
Supporting research data are available on reasonable request from the corresponding author.

\bibliographystyle{apsrev4-1}
\bibliography{frb_bib.bib}

\end{document}